\documentclass[aps,prd,a4paper,amsmath,amssymb,nofootinbib,showpacs,showkeys,
preprintnumbers,notitlepage,superscriptaddress,twocolumn]{revtex4-1}

\usepackage{bm}
\usepackage[dvips]{color,graphicx}

\begin{document}

\title{Magnetic Susceptibility of the Quark Condensate and
Polarization from Chiral Models}

\author{Marco Frasca}\email{marcofrasca@mclink.it}
\affiliation{via Erasmo Gattamelata, 3, 00176 Roma, Italy}
\author{Marco Ruggieri}\email{ruggieri@yukawa.kyoto-u.ac.jp}
\affiliation{Yukawa Institute for Theoretical Physics, Kyoto
University, Kitashirakawa Oiwake-cho, Sakyo-ku, Kyoto 606-8502,
Japan}



\begin{abstract}
We compute the magnetic susceptibility of the quark condensate and
the polarization of quarks at zero temperature and in a uniform
magnetic background. Our theoretical framework consists of two
chiral models that allow to treat self-consistently the
spontaneous breaking of chiral symmetry: the linear $\sigma-$model
coupled to quarks, dubbed quark-meson model, and the
Nambu-Jona-Lasinio model. We also perform analytic estimates of
the same quantities within the renormalized quark-meson model,
both in the regimes of weak and strong fields. Our numerical
results are in agreement with the recent literature; moreover, we
confirm previous Lattice findings, related to the saturation of
the polarization at large fields.
\end{abstract}

 \preprint{YITP-11-28}\maketitle


\section{Introduction}
One of the most attractive aspects of the vacuum of Quantum
Chromodynamics (QCD), the theory of strong interactions, is its
non-perturbative nature. Mainly by means of Lattice QCD
simulations at zero quark chemical potential
~\cite{deForcrand:2006pv,Aoki:2009sc,Bazavov:2009zn,Cheng:2009be,Karsch:2000kv}
it is established that two crossovers take place in a narrow range
of temperature; one for quark deconfinement, and another one for
the (approximate) restoration of chiral symmetry. Besides,
powerful analytic and semi-analytic techniques have been developed
to understand the coupling between chiral symmetry restoration and
deconfinement, see~\cite{Fischer:2009wc,Frasca:2007uz} and
references therein. Moreover, Lattice QCD and Operator Product
Expansion (OPE in the following) of the correlators of hadronic
currents show that the QCD vacuum can be characterized by several
quark, gluon and mixed condensates~\cite{Shifman:1978bx}.

A fruitful theoretical approach to the physics of strong
interactions, which is capable to capture some of the
non-perturbative properties of the QCD vacuum, is the use of
effective chiral models. One of them is the Nambu-Jona-Lasinio
(NJL) model~\cite{Nambu:1961tp} (see Refs.~\cite{revNJL} for
reviews), in which the QCD gluon-mediated interactions are
replaced by effective interactions among quarks, which are built
in order to respect the global symmetries of QCD. Under some
approximations, it is possible to derive the NJL model effective
interaction kernel from first principles QCD,
see~\cite{Kondo:2010ts,Frasca:2010iv}. Besides, a linear
$\sigma-$model augmented with a Yukawa-type coupling to quarks,
named Quark-Meson model (QM model in the following), has been
developed as an effective model of
QCD~\cite{Gervais:1969zz,Mota:1999hb}. In this model, quartic
meson self-couplings allow to absorb the cutoff dependence of the
coupling constants; moreover, following an idea by
Weinberg~\cite{Weinberg:1997rv}, tree level propagating mesons are
sufficient to avoid triviality of the NJL model in $3+1$
dimensions in the one-loop approximation~\cite{Wilson:1972cf}
(see~\cite{Mota:1999hb} for an excellent discussion about these
points).

The chiral models are widely used to map qualitatively, and to
some extent also quantitatively, the phase diagram of strongly
interacting matter along several directions like temperature,
chemical potential, isospin chemical potential and external
fields~\cite{Roessner:2006xn,Fukushima:2008wg,Abuki:2008nm,
Kashiwa:2007hw,Herbst:2010rf,Skokov:2010sf,Kahara:2008yg,Fukushima:2010fe,
Mizher:2010zb,Gatto:2010pt,Suganuma:1990nn,
Fraga:2008qn,Campanelli:2009sc,Chernodub:2011fr}. It is thus
interesting to use these models to compute other quantities, which
are related to the QCD vacuum condensates. As a matter of fact,
the chiral models allow for a self-consistent treatment of the
spontaneous chiral symmetry breaking in the vacuum. Therefore,
after the ground state properties (i.e., the chiral condensate in
the NJL model, or the expectation value of the $\sigma$-field in
the quark-meson model) under external factors are computed, it is
straightforward to estimate numerical values of other condensates,
using the one-loop propagators of the theory.

It has been realized that external fields can induce QCD
condensates that are absent otherwise~\cite{Ioffe:1983ju}. Of
particular interest for this article is magnetic moment,
$\langle\bar f \Sigma^{\mu\nu} f\rangle$ where $f$ denotes the
fermion field of the flavor $f-$th, and $\Sigma^{\mu\nu} =
-i(\gamma^\mu \gamma^\nu - \gamma^\nu\gamma^\mu)/2$. At small
fields one can write, according to~\cite{Ioffe:1983ju},
\begin{equation}
\langle\bar f \Sigma^{\mu\nu} f\rangle = \chi\langle\bar f
f\rangle Q_f |eB|~, \label{eq:chiDef}
\end{equation}
and $\chi$ is a constant independent on flavor, which is dubbed
magnetic susceptibility of the quark condensate.
In~\cite{Ioffe:1983ju} it is proved that the role of the
condensate~\eqref{eq:chiDef} to QCD sum rules in external fields
is significant, and it cannot be ignored. The quantity $\chi$ has
been computed by means of special sum
rules~\cite{Ioffe:1983ju,Belyaev:1984ic,Balitsky:1985aq,Ball:2002ps,Rohrwild:2007yt},
OPE combined with Pion Dominance~\cite{Vainshtein:2002nv},
holography~\cite{Gorsky:2009ma,Son:2010vc}, instanton vacuum
model~\cite{Kim:2004hd}, analytically from the zero mode of the
Dirac operator in the background of a $SU(2)$
instanton~\cite{Ioffe:2009yi}, and on the Lattice in two color
quenched simulations at zero and finite
temperature~\cite{Buividovich:2009ih}. It has also been suggested
that in the photoproduction of lepton pairs, the interference of
the Drell-Yan amplitude with the amplitude of a process where the
photon couples to quarks through its chiral-odd distribution
amplitude, which is normalized to the magnetic susceptibility of
the QCD vacuum, is possible~\cite{Pire:2009ap}. This interference
allows in principle to access the chiral odd transversity parton
distribution in the proton. Therefore, this quantity is
interesting both theoretically and phenomenologically. The several
estimates, that we briefly review in Section III, lead to the
numerical value of $\chi$ as follows:
\begin{equation}
\chi\langle\bar ff\rangle = 40-70~\text{MeV}~. \label{eq:i1}
\end{equation}

A second quantity, which embeds non-linear effects at large
fields, is the polarization, $\mu_f$, defined as
\begin{equation}
\mu_f = \left|\frac{\Sigma_f}{\langle\bar f f\rangle}\right|~,~~~
\Sigma_f = \langle\bar f\Sigma^{12}f\rangle~, \label{eq:muDef}
\end{equation}
which has been computed on the Lattice
in~\cite{Buividovich:2009ih} for a wide range of magnetic fields,
in the framework of two-color QCD with quenched fermions. At small
fields $\mu_f = |\chi Q_f eB|$ naturally; at large fields,
non-linear effects dominate and an interesting saturation of
$\mu_f$ to the asymptotic value $\mu_\infty = 1$ is measured.
According to~\cite{Buividovich:2009ih} the behavior of the
polarization as a function of $eB$ in the whole range examined,
can be described by a simple inverse tangent function. Besides,
magnetization of the QCD vacuum has been computed in the strong
field limit in~\cite{Cohen:2008bk} using perturbative QCD, where
it is found it grows as $B\log B$.

In this article, we compute the magnetic susceptibility of the
quark condensate by means of the NJL and the QM models. This study
is interesting because in the chiral models, it is possible to
compute self-consistently the numerical values of the condensates
as a function of $eB$, once the parameters are fixed to reproduce
some characteristic of the QCD vacuum. We firstly perform a
numerical study of the problem, which is then complemented by some
analytic estimate of the same quantity within the renormalized QM
model. Moreover, we compute the polarization of quarks at small as
well as large fields, both numerically and analytically. In
agreement with the Lattice results~\cite{Buividovich:2009ih}, we
also measure a saturation of $\mu_f$ to one at large fields, in
the case of the effective models. Our results push towards the
interpretation of the saturation as a non-artifact of the Lattice.
On the contrary, we can offer a simple physical understanding of
this behavior, in terms of lowest Landau level dominance of the
chiral condensate. As a matter of fact, using the simple equations
of the models for the chiral condensate and for the magnetic
moment, we can show that at large magnetic field $\mu_f$ has to
saturate to one, because in this limit the higher Landau levels
are expelled from the chiral condensate; as a consequence, the
ratio of the two approaches one asymptotically.

We also obtain a saturation of the polarization within the
renormalized QM model. There are some differences, however, in
comparison with the results of the non-renormalized models. In the
former case, the asymptotic value of $\mu_f$ is charge-dependent;
moreover, the interpretation of the saturation as a lowest Landau
level (LLL) dominance is not straightforward, because the
renormalized contribution of the higher Landau levels is important
in the chiral condensate, even in the limit of very strong fields.
It is possible that the results obtained within the renormalized
model are a little bit far from true QCD. As a matter of fact, in
the renormalized model we assume that the quark self-energy is
independent on momentum; thus, when we take the limit of infinite
quark momentum in the gap equation, and absorb the ultraviolet
divergences by means of counterterms and renormalization
conditions, we implicitly assume that that quark mass at large
momenta is equal to its value at zero momentum. We know that this
is not true, see for
example~\cite{Politzer:1976tv,Langfeld:1996rn}: even in the
renormalized theory, the quark self-energy naturally cuts off the
large momenta, leading to LLL dominance in the traces of quark
propagator which are relevant for our study. Nevertheless, it is
worth to study this problem within the renormalized QM model in
its simplest version, because it helps to understand the structure
of this theory under the influence of a strong magnetic field.

In our calculations we neglect, for simplicity, the possible
condensation of $\rho-$mesons at strong
fields~\cite{Chernodub:2010qx,Chernodub:2011mc}. Vector meson
dominance~\cite{Chernodub:2010qx} and Sakai-Sugimoto
model~\cite{Callebaut:2011uc} suggest for the condensation a
critical value of $eB_c \approx m_\rho^2\approx 0.57$ GeV$^2$,
where $m_\rho$ is the $\rho-$meson mass in the vacuum. Beside
these, a NJL-based calculation within the lowest Landau level
(LLL) approximation~\cite{Chernodub:2011mc} predicts $\rho-$meson
condensation at strong fields as well, even if in the latter case
it is hard to estimate exactly $e B_c$, mainly because of the
uncertainty of the parameters of the model. It would certainly be
interesting to address this problem within our calculations, in
which not only the LLL but also the higher Landau levels are
considered, and in which the spontaneous breaking of chiral
symmetry is kept into account self-consistently. However, this
would complicate significantly the calculational setup. Therefore,
for simplicity we leave this issue to a future project.

The plan of the article is as follows. In Section II we describe
the QM and NJL models and fix our notation. In Section III we
discuss our numerical results for the polarization of quarks, and
compute the magnetic susceptibility of the quark condensate. In
Section IV we compute the renormalized Quantum Effective Potential
(QEP) of the QM model in a magnetic background, in the one-loop
approximation, and compute analytically the solution of the gap
equation in the weak field case, and semi-analytically in the
strong field limit. We then use the results to estimate $\mu_f$
and $\chi$. Finally, in Section V we draw our conclusions.

\section{Chiral models coupled to a magnetic field}
In the first part of this article, we derive numerical results for
the spin polarization of quarks in a magnetic field, and for the
magnetic susceptibility of the quark condensate, using two
effective chiral models: the Nambu-Jona-Lasinio (NJL) model, and
the Quark-Meson (QM) model. We describe the two models in some
detail in this Section. We work in the Landau gauge and take the
magnetic field along the $z-$axis, $\bm B = (0,0,B)$.

\subsection{Quark-Meson model}
In the QM model, a meson sector described by the linear sigma
model lagrangian, is coupled to quarks via a Yukawa-type
interaction. The model is renormalizable in $D=3+1$ dimensions.
However, since we adopt the point of view of it as an effective
description of QCD, it is not necessary to use the renormalized
version of the model itself. On the contrary, it is enough to fix
an ultraviolet scale to cutoff the divergent expectation values;
the UV scale is then chosen phenomenologically, by requiring that
the numerical value of the chiral condensate in the vacuum
obtained within the model, is consistent with the results obtained
from the sum rules~\cite{Dosch:1997wb}. This is a rough
approximation of the QCD effective quark mass, which smoothly
decays at large momenta~\cite{Langfeld:1996rn,Politzer:1976tv}. In
Section IV we will use a renormalized version of the model, to
derive semi-analytically some results in the two regimes of weak
and strong fields.

The lagrangian density of the model is given by
\begin{eqnarray}
{\cal L} &=& \bar q \left[iD_\mu\gamma^\mu - g(\sigma +
i\gamma_5\bm\tau\cdot\bm\pi)\right] q
\nonumber \\
&& + \frac{1}{2}\left(\partial_\mu\sigma\right)^2 +
\frac{1}{2}\left(\partial_\mu\bm\pi\right)^2 - U(\sigma,\bm\pi)~.
\label{eq:LD1}
\end{eqnarray}
In the above equation, $q$ corresponds to a quark field in the
fundamental representation of color group $SU(3)$ and flavor group
$SU(2)$; the covariant derivative, $D_\mu = \partial_\mu - Q_f e
A_\mu$, describes the coupling to the background magnetic field,
where $Q_f$ denotes the charge of the flavor $f$. Besides,
$\sigma$, $\bm\pi$ correspond to the scalar singlet and the
pseudo-scalar iso-triplet fields, respectively. The potential $U$
describes tree-level interactions among the meson fields. In this
article, we take its analytic form as
\begin{equation}
U(\sigma,\bm\pi) = \frac{\lambda}{4}\left(\sigma^2
+\bm\pi^2-v^2\right)^2 - h\sigma~, \label{eq:U}
\end{equation}
where the first addendum is chiral invariant; the second one
describes a soft explicit breaking of chiral symmetry, and it is
thus responsible for the non-zero value of the pion mass. For
$h=0$, the interaction terms of the model are invariant under
$SU(2)_V\otimes SU(2)_A\otimes U(1)_V$. This group is broken
explicitly to $U(1)_V^3\otimes U(1)_A^3\otimes U(1)_V$ if the
magnetic field is coupled to the quarks, because of the different
electric charge of $u$ and $d$ quarks. Here, the superscript $3$
in the $V$ and $A$ groups denotes the transformations generated by
$\tau_3$, $\tau_3\gamma_5$ respectively. Therefore, the chiral
group in presence of a magnetic field is $U(1)_V^3\otimes
U(1)_A^3$. This group is then explicitly broken by $h$-term to
$U(1)_V^3$.

In this article, we restrict ourselves to the large-$N_c$ (that
is, one-loop) approximation, which amounts to consider mesons as
classical fields, and integrate only over fermions in the
generating functional of the theory to obtain the Quantum
Effective Potential (QEP). As a matter of fact, quantum
corrections arising from meson bubbles are suppressed of a factor
$1/N_c$ with respect to case of the fermion bubble. In the
integration process, the meson fields are fixed to their classical
expectation value, $\langle\bm\pi\rangle = 0$ and
$\langle\sigma\rangle \neq 0$ (in particular, $\sigma$ has the
quantum numbers of the chiral condensate, $\langle\bar q
q\rangle$). The physical value of $\langle\sigma\rangle$ will be
then determined by minimization of the QEP.

To compute QEP in presence of a magnetic background, we use the
Leung-Ritus-Wang method~\cite{Ritus:1972ky} which allows to write
down the quark propagator for the flavor $f$ in terms of Landau
levels,
\begin{equation}
S_f(x,y) = \sum_{k=0}^\infty\int\frac{dp_0 dp_2 dp_3}{(2\pi)^4}
E_P(x)\Lambda_k \frac{i}{P\cdot\gamma - M}\bar{E}_P(y)~,
\label{eq:QP}
\end{equation}
where $E_P(x)$ corresponds to the eigenfunction of a charged
fermion in magnetic field, and $\bar{E}_P(x) \equiv
\gamma_0(E_P(x))^\dagger \gamma_0$. In the above equation,
\begin{equation}
P = (p_0,0,{\cal Q}\sqrt{2k|Q_f eB|},p_3)~,\label{eq:MB}
\end{equation}
where $k =0,1,2,\dots$ labels the $k^{\text{th}}$ Landau level,
and ${\cal Q} \equiv\text{sign}(Q_f)$, with $Q_f$ denoting the
charge of the flavor $f$; $\Lambda_k$ is a projector in Dirac
space which keeps into account the degeneracy of the Landau
levels; it is given by
\begin{equation}
\Lambda_k = \delta_{k0}\left[{\cal P_+}\delta_{{\cal Q},+1} +
{\cal P_-}\delta_{{\cal Q},-1}\right] + (1-\delta_{k0})I~,
\end{equation}
where ${\cal P}_{\pm}$ are spin projectors and $I$ is the identity
matrix in Dirac spinor indices. At the one-loop level, the QEP
then reads
\begin{eqnarray}
V &=& \frac{\lambda}{4}\left(\sigma^2-v^2\right)^2
- h\sigma \nonumber \\
&& -N_c\sum_f\frac{|Q_f
eB|}{2\pi}\sum_k\beta_k\int\frac{dp_3}{2\pi}\omega_k(p_3)~.\label{eq:QEA}
\end{eqnarray}
In the above equation we have defined
\begin{equation}
\omega_k(p_3) = \sqrt{p_3^2 + 2 k |Q_f eB| + m_q^2}~,
\label{eq:1ps}
\end{equation}
with $m_q = g\sigma$; $\beta_k = 2-\delta_{k0}$ counts the
degeneracy of the Landau levels.

The one-loop fermion contribution, which corresponds to the last
addendum in the r.h.s. of Eq.~\eqref{eq:QEA}, is divergent in the
ultraviolet. In order to regularize it, we adopt a smooth
regulator $U_\Lambda$ as in~\cite{Gatto:2010pt}, which is more
suitable, from the numerical point of view, in our model
calculation with respect to the hard-cutoff which is used in
analogous calculations without magnetic field. In this article we
chose
\begin{equation}
U_\Lambda = \frac{\Lambda^{2N}}{\Lambda^{2N} + (p_z^2 + 2|Q_f e
B|k)^N}~,~~~~~N=5~. \label{eq:UV}
\end{equation}
The (more usual) 3-momentum cutoff regularization scheme is
recovered in the limit $N\rightarrow\infty$; we notice that, even
if the choice $N=5$ may seem arbitrary to some extent, it is not
more arbitrary than the choice of the hard cutoff scheme, that is,
of a regularization scheme. In effective models, the choice of a
regularization scheme is a part of the definition of the model
itself. Momentum integrals are understood as follows:
\begin{equation}
\sum_n\beta_n\int\frac{dp}{2\pi} \rightarrow
\sum_n\beta_n\int\frac{dp}{2\pi}U_\Lambda~. \label{eq:repla}
\end{equation}

Once the expectation value of $\sigma$ is computed as a function
of $eB$ by a procedure of minimization of $V$, we compute the
expectation values that are relevant to the context. To begin
with, we consider the chiral condensate for the flavor $f$,
$\langle\bar f f\rangle = -\text{Tr}[S_f(x,x)]$, with $S_f$ given
by Eq.~\eqref{eq:QP}. It is straightforward to derive the relation
\begin{equation}
\langle\bar f f\rangle = - N_c\frac{|Q_f
eB|}{2\pi}\sum_{k=0}^\infty \beta_k \int\frac{dp_3}{2\pi}
\frac{m_q}{\omega_k(p_3) }~,\label{eq:CC}
\end{equation}
where the divergent integral on the r.h.s. of the above equation
has to be understood regularized as in~\eqref{eq:repla}. From
Eq.~\eqref{eq:CC} we notice that the prescription~\eqref{eq:repla}
is almost equivalent to the introduction of a running effective
quark mass,
\begin{equation}
m_q = g\sigma\Theta\left(\Lambda^2 - p_3^2 - 2k|Q_f eB|\right)~,
\label{eq:RM}
\end{equation}
that can be considered as a rough approximation to the effective
running quark mass in QCD~\cite{Politzer:1976tv} which decays at
large quark momenta, see also the discussion
in~\cite{Langfeld:1996rn}. Once the scale $\Lambda$ is fixed, the
Landau levels with $n\geq 1$ are removed from the chiral
condensate if $eB \gg \Lambda^2$.

Next we turn to the magnetic moment for the flavor $f$,
\begin{equation}
\langle \bar f \Sigma^{\mu\nu} f\rangle =
-\text{Tr}[\Sigma^{\mu\nu}S_f(x,x)]~,\label{eq:MDef}
\end{equation}
where
\begin{equation}
\Sigma^{\mu\nu} = \frac{1}{2i}(\gamma^\mu \gamma^\nu -
\gamma^\nu\gamma^\mu)~,
\end{equation}
is the relativistic spin operator. We take $\bm B = (0,0,B)$; in
this case, only $\Sigma^{12}\equiv\Sigma_f$ is non-vanishing.
Using the properties of $\gamma-$matrices it is easy to show that
only the Lowest Landau Level (LLL) gives a non-vanishing
contribution to the trace:
\begin{equation}
\Sigma_f = N_c \frac{Q_f |eB|}{2\pi}\int\frac{dp_3}{2\pi}
\frac{m_q}{\omega_0(p_3) }~,\label{eq:TrM}
\end{equation}
where $\omega_0 = \omega_{k=0}$. Once again, the divergent
integral on the r.h.s. of the above equation has to be understood
regularized following the prescription in Eq.~\eqref{eq:repla}.

For the QM model, the parameters are tuned as follows. We fix the
constituent quark mass in the vacuum to a phenomenological value,
$m_q = 335$ MeV; furthermore, we require that in the vacuum the
following condition holds:
\begin{equation}
\left.\frac{\partial V(\sigma,\bm
B=0)}{\partial\sigma}\right|_{\sigma=f_\pi} = 0~, \label{eq:EVv}
\end{equation}
which implies that $\langle\sigma\rangle = f_\pi$ in the vacuum,
thus $m_q = g f_\pi$. As a consequence we find $g=3.62$. The
parameter $h$ is fixed by the condition $h=f_\pi m_\pi^2$, where
$m_\pi$ is the average value of the pion mass in the vacuum and
$f_\pi = 92.4$ MeV. We have then $h=0.047 m_q^3$. To determine $v$
and $\lambda$ we solve simultaneously Eq.~\eqref{eq:EVv} and
\begin{equation}
m_\sigma^2 = \left.\frac{\partial^2
V}{\partial\sigma^2}\right|_{\sigma = f_\pi}~, \label{eq:msigma}
\end{equation}
with $m_\sigma = 700$ MeV. The divergences in these two equations
are cured with the prescription~\eqref{eq:repla}. In the UV
regulator we chose $\Lambda=560$ MeV which implies $\langle\bar
uu\rangle = (-231~\text{MeV})^3$. This results in the values
$\lambda = 4.67$ and $v^2 = -1.8 m_q^2$. Before going ahead, we
notice that in the non-renormalized model, the 1-loop fermion
contribution regularized at the scale $\Lambda$ is included into
the conditions~\eqref{eq:EVv} and~\eqref{eq:msigma}.

\subsection{Nambu-Jona-Lasinio model}
The quark lagrangian density of the NJL model is given by
\begin{equation}
{\cal L} =\bar q\left(i\gamma^\mu D_\mu - m_0\right) q +
G\left[\left(\bar qq\right)^2 + \left(i\bar q\gamma_5\bm\tau
q\right)^2\right]~;\label{eq:1ooo}
\end{equation}
here $q$ is the quark Dirac spinor in the fundamental
representation of the flavor $SU(2)$ and the color group;
$\bm\tau$ correspond to the Pauli matrices in flavor space. A sum
over color and flavor is understood. Once again, the covariant
derivative embeds the QED coupling of the quarks with the external
magnetic field. The explicit soft breaking of chiral symmetry in
this model is achieved by introducing a current quark mass, $m_0$,
whose numerical value is fixed by fitting the vacuum pion mass.
The NJL model is not renormalizable in the usual sense for $D>2$
space-time dimensions; it is renormalizable in $D<4$ in the
one-loop approximation. At $D=4$ it might represent a trivial
theory of non-interacting bosons if the ultraviolet cutoff is
allowed to be infinite~\cite{Mota:1999hb,Wilson:1972cf}. However,
adopting the point of view of the NJL model as an effective (and
rough) description of low-energy QCD, it is not necessary to
require the cutoff to be infinite: the cutoff can be interpreted
as a physical quantity which makes the effective quark mass to be
constant at small quark momenta, and vanishing at large momenta,
thus mimicking roughly the running effective mass of real
QCD~\cite{Politzer:1976tv}. It is worth to notice that a non-local
interaction among quarks with the same quantum numbers of the
four-fermion local term in Eq.~\eqref{eq:1ooo} can be derived by
the QCD action under some approximation, taking the low energy
limit of the latter~\cite{Kondo:2010ts,Frasca:2007uz}. For
simplicity, we treat here only the case of local interaction.

Once again, the one-loop fermion contribution can be obtained
within the Leung-Ritus-Wang method:
\begin{equation}
V = G\Sigma^2  -N_c\sum_{f=u,d}\frac{|Q_f
eB|}{2\pi}\sum_{n}\beta_n\int_{-\infty}^{+\infty}\frac{dp }{2\pi}
\omega_n(p)~,\label{eq:OB}
\end{equation}
with $\omega_n(p) = \sqrt{p^2 + 2 n |Q_f eB| + M_q^2}$, and $M_q =
m_0 -2G\Sigma$ with $\Sigma = \langle\bar u u + \bar d d\rangle$.
Equations~\eqref{eq:CC} and~\eqref{eq:MDef} are still valid in the
NJL model, with the replacement $m_q \rightarrow M_q$. A
comparison of the QEPs of the QM model and of the NJL model shows
that at the one-loop level, the two models differ only for the
classic part of the effective meson potential, and for the
definition of the constituent quark mass.

The parameters in the NJL model are fixed as follows. The bare
quark mass is computed by virtue of the Gell-Mann-Oakes-Renner
relation, $m_\pi^2 f_\pi^2= -2m_0 \langle\bar uu\rangle$, which is
satisfied in the NJL model~\cite{revNJL}. Then, two equations are
solved simultaneously: one for $f_\pi$ and one for the chiral
condensate in the vacuum at zero magnetic field strength.
Requiring that $f_\pi = 92.4$ MeV and $\langle\bar uu\rangle =
(-253~\text{MeV})^3$, this procedure gives the numerical values of
$G$ and $\Lambda$. Thus we find $\Lambda = 626.76$ MeV, $G
=2.02/\Lambda^2$ and $m_0 = 5$ MeV.

\section{Numerical results}

In this Section, we collect our numerical results which are
relevant for the computation of the magnetic susceptibility of the
chiral condensate, and of the spin polarization. From the
numerical point of view, it is more convenient to compute firstly
the latter; then, a fit of the polarization data at small fields
will enable to extract the value of the magnetic susceptibility of
the quark condensate. For what concerns data about chiral
condensate, magnetic moment and polarization, we plot results only
for the QM model, since the results obtained within the NJL model
are qualitatively very similar to those obtained within the former
model. Then, we will give the final result for the chiral
magnetization for the two models considered here.

\subsection{Polarization}

\begin{figure}[t!]
\begin{center}
\includegraphics[width=8.5cm]{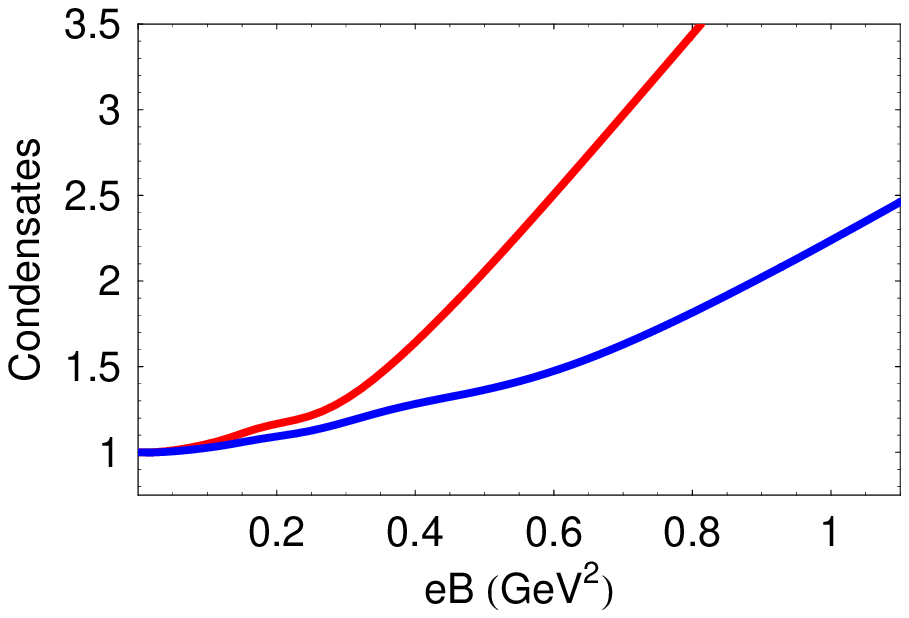}\\
\includegraphics[width=8.5cm]{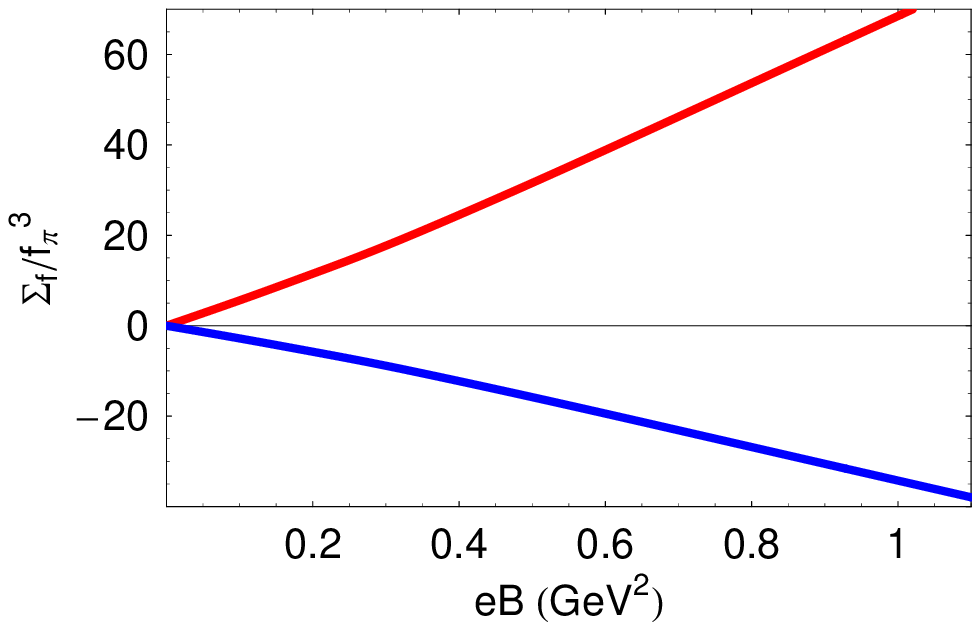}
\end{center}
\caption{\label{Fig:1} {\em Upper panel.} Chiral condensates of
$u$-quarks (red) and $d$-quarks (blue), in units of the same
quantities at zero magnetic field, as a function of the magnetic
field.  {\em Lower panel.} Expectation value of the magnetic
moment operator, in units of $f_\pi^3$, as a function of $eB$.
Data correspond to the QM model.}
\end{figure}

The physical value of the (total) chiral condensate $\Sigma$ for
the NJL model, and of $\sigma$ for the QM model, are obtained
numerically by a minimization procedure of the QEP~\eqref{eq:OB}
for any value of $eB$. Then, we make use of Eqs.~\eqref{eq:CC}
and~\eqref{eq:MDef}, with the replacement~\eqref{eq:repla}, to
compute the chiral condensate and the magnetic moment for each
flavor.

In the upper panel of Fig.~\ref{Fig:1}, we plot the chiral
condensates for $u$ and $d$ quarks, as a function of $eB$, for the
QM model. The magnetic field splits the two quantities because of
the different charge for the two quarks. The small oscillations,
which are more evident for the case of the $u$-quark, are an
artifact of the regularization scheme, and disappear if smoother
regulators are used, see the discussion
in~\cite{Campanelli:2009sc}. In the regime of weak fields, our
data are consistent with the scaling $\langle\bar ff\rangle
\propto |eB|^2/M$ where $M$ denotes some mass scale; in the strong
field limit we find instead $\langle\bar ff\rangle \propto
|eB|^{3/2}$. The behavior of the quark condensate as a function of
magnetic field is in agreement with the magnetic catalysis
scenario~\cite{Suganuma:1990nn,Klevansky:1989vi}.

In the lower panel of Fig.~\ref{Fig:1} we plot our data for the
expectation value of the magnetic moment. Data correspond to the
QM model (for the NJL model we obtain similar results). At weak
fields, $\Sigma_f \propto |eB|$ as expected from
Eq.~\eqref{eq:MDef}. In the strong field limit, non-linearity
arise because of the scaling of quark mass (or chiral condensate);
we find $\Sigma_f \propto |eB|^{3/2}$ in this limit.

\begin{figure}[t!]
\begin{center}
\includegraphics[width=8.5cm]{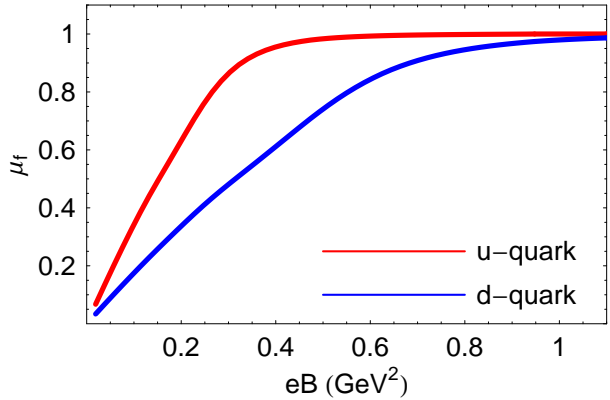}
\end{center}
\caption{\label{Fig:2} Polarization of $u$-quarks (red) and
$d-$quarks (blue) as a function the magnetic field strength, for
the QM model. }
\end{figure}

In Fig.~\ref{Fig:2} we plot our results for the polarization. Data
are obtained by the previous ones, using the
definition~\eqref{eq:muDef}. At small fields, the polarization
clearly grows linearly with the magnetic field. This is a natural
consequence of the linear behavior of the magnetic moment as a
function of $eB$ for small fields, see Fig.~\ref{Fig:1}. On the
other hand, within the chiral models we measure a saturation of
$\mu_f$ at large values of $eB$, to an asymptotic value
$\mu_\infty = 1$. This conclusion remains unchanged if we consider
the NJL model, and it is in agreement with the recent Lattice
findings~\cite{Buividovich:2009ih}. It should be noticed that, at
least for the $u-$quark, saturation is achieved before the
expected threshold for $\rho-$meson
condensation~\cite{Chernodub:2010qx,Chernodub:2011mc,Callebaut:2011uc}.
Therefore, our expectation is that our result is stable also if
vector meson condensation is considered.

The saturation to the asymptotic value $\mu_\infty = 1$ of
polarization is naturally understood within the models we
investigate, as a LLL dominance in the chiral condensate (i.e.,
full polarization). As a matter of fact, $\Sigma_f$ and
$\langle\bar ff\rangle$ turn to be proportional in the strong
field limit, since only the LLL gives a contribution to the the
latter, compare Eq.~\eqref{eq:CC} and~\eqref{eq:TrM} which imply
\begin{equation}
\mu_f = 1 - \frac{\langle\bar ff\rangle_{\text{HLL}}}{\langle\bar
ff\rangle}~, \label{eq:HLL}
\end{equation}
where $\langle\bar ff\rangle_{\text{HLL}}$ corresponds to the
higher Landau levels contribution to the chiral condensate. In the
strong field limit $\langle\bar ff\rangle_{\text{HLL}} \rightarrow
0$ because of Eq.~\eqref{eq:repla} which is a rough approximation
to the QCD effective quark mass, as depicted in Eq.~\eqref{eq:RM};
hence, $\mu_f$ has to approach the asymptotic value $\mu_\infty =
1$.  On the other hand, in the weak field limit $\langle\bar
ff\rangle_{\text{HLL}} \rightarrow \langle\bar ff\rangle$ and the
proportionality among $\Sigma_f$ and $\langle\bar ff\rangle$ is
lost.

\begin{figure}[t!]
\begin{center}
\includegraphics[width=8.5cm]{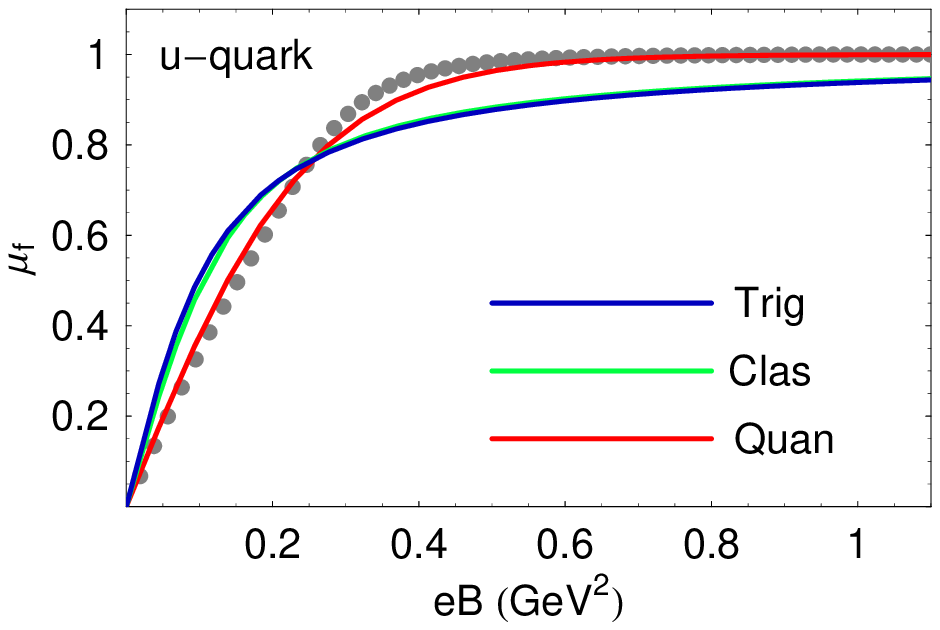}\\
\includegraphics[width=8.5cm]{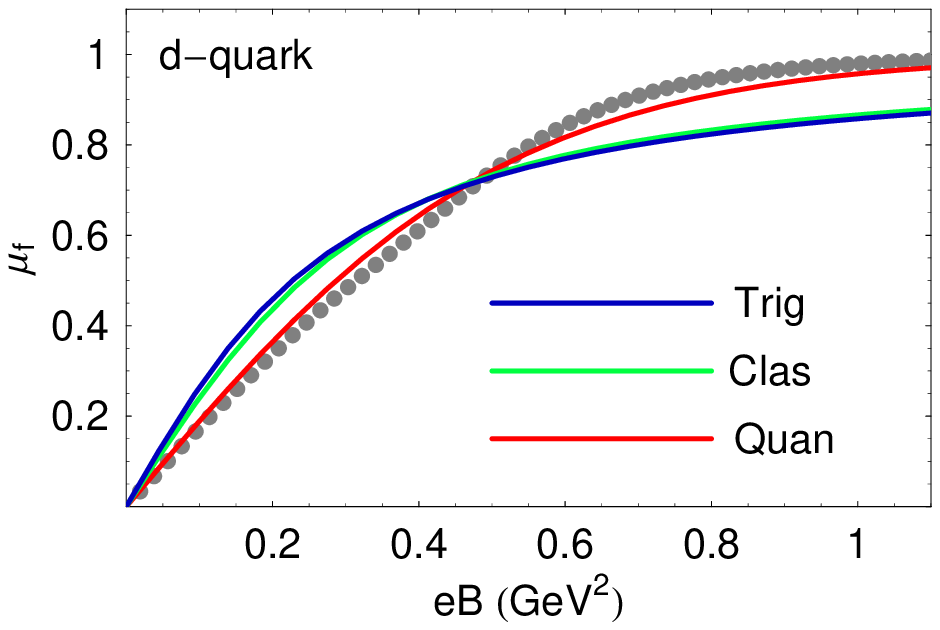}
\end{center}
\caption{\label{Fig:fak} {\em Upper panel.} Polarization data for
$u-$quark (grey dots) and fitting curves. See the text for
details. {\em Lower panel.} Polarization data and fitting
functions for $d-$quark. Results correspond to the QM model.}
\end{figure}

In Ref.~\cite{Buividovich:2009ih}, data of polarization are fit by
means of three different functions, namely
\begin{eqnarray}
\mu^{\text{clas}}_f &=& \mu_\infty \left|\coth\frac{3\chi Q_f eB
}{\mu_\infty}   - \frac{\mu_\infty}{3\chi Q_f eB }\right|~, \label{eq:CLas}\\
\mu^{\text{quan}}_f &=& \mu_\infty \left|2\coth\frac{2\chi Q_f eB
}{\mu_\infty}   - \coth\frac{\chi Q_f eB }{\mu_\infty}\right|~, \label{eq:QUan}\\
\mu^{\text{trig}}_f &=&
\frac{2\mu_\infty}{\pi}\arctan\left|\frac{\pi\chi Q_f
eB}{2\mu_\infty}\right| \label{eq:TRig}~.
\end{eqnarray}
The three functions above share the behavior at the origin,
$\mu_{f}\approx |\chi Q_f eB|$, and the asymptotic one, $\mu_f
\approx \mu_\infty$. In~\cite{Buividovich:2009ih} a two-parameter
fit on $\chi$ and $\mu_\infty$ is performed; the model with the
lowest chi-squared per degrees of freedom is represented by the
trigonometric one, Eq.~\eqref{eq:TRig}. Inspired by these results
we have tried to fit our data using the same
functions~\eqref{eq:CLas}-\eqref{eq:TRig}. In our model
calculation, the asymptotic value $\mu_\infty = 1$ is achieved
straightforwardly, therefore it is enough to perform a
one-parameter fit leaving $\chi$ as a free parameter. The results
of this procedure are collected in Fig.~\ref{Fig:fak}, where we
plot our data of quark's polarization as a function of $eB$ (gray
dots) and the three fitting functions with $\mu_\infty = 1$. From
Fig.~\ref{Fig:fak} we read that both the trigonometric and the
classic functions do not adapt well to our data (for the NJL model
we obtain similar results). In particular, both of the
aforementioned functions overestimate $\chi$, and reach slowly the
asymptotic value. On the other hand, the quantum function in
Eq.~\eqref{eq:QUan} offers a better description of our data in the
whole range of $eB$ examined here. We have checked that the value
of $\chi$ obtained within this fit overestimates the value
obtained by a weak-field linear fit only of the $10\%$, see below.
Moreover, the fit functions smoothly follow the large $eB$ data to
the asymptotic value $\mu_f = 1$. We conclude that within our
model, the quantum fitting function is a more faithful
representation of data. The difference with Lattice data is
probably due to the fact that the latter ones are obtained with
quenched (hence, non-dynamical) fermions, while in our case
dynamical fermions are considered. It will be interesting,
therefore, to compare the model predictions with Lattice data
obtained with other kinds of fermions in the near future.

Before going ahead, it is interesting that our data on
polarization, and our interpretation of the saturation of the
latter at large fields, gives a quantitative estimate of the
goodness of the LLL approximation for the models at hand. In
particular, at $eB_c \approx 0.6$ GeV$^2$, which in turns is the
estimated critical value for vector meson
condensation~\cite{Chernodub:2010qx,Callebaut:2011uc}, the LLL
almost dominates the chiral condensate of the $u-$quark; on the
other hand, at this value of $eB$ the LLL dominance has not yet
been achieved for the $d-$quark condensate; but the higher Landau
levels at this value of $eB$ give a contribution almost to the
$20\%$ of the chiral condensate. This is an indirect check of the
consistence of the LLL approximation used
in~\cite{Chernodub:2011mc}, which in turn should be good within a
$20\%$ accuracy.

\subsection{Magnetic susceptibility of the quark condensate}

At small fields $\mu_f = |\chi Q_f eB|$ from
Eq.~\eqref{eq:chiDef}. Hence, we use the data on polarization at
small fields, to obtain the numerical value of the magnetic
susceptibility of the chiral condensate. Our results are as
follows:
\begin{eqnarray}
\chi &\approx& -4.3~\text{GeV}^{-2}~,~~~\text{NJL}
\\
\chi &\approx& -5.25~\text{GeV}^{-2}~,~~~\text{QM}
\end{eqnarray}
respectively for the NJL model and the QM model. To obtain the
numerical values above we have used data for $eB$ up to $5 m_\pi^2
\approx 0.1$ GeV$^2$, which are then fit using a linear law. Using
the numerical values of the chiral condensate in the two models,
we obtain
\begin{eqnarray}
\chi \langle\bar f f \rangle &\approx&
69~\text{MeV}~,~~~\text{NJL} \\
\chi \langle\bar f f \rangle &\approx& 65~\text{MeV}~,~~~\text{QM}
\end{eqnarray}
The numerical values of $\chi$ that we obtain within the effective
models are in fair agreement with recent results, see Table I. In
our model calculations, the role of the renormalization scale is
played approximately by the ultraviolet cutoff, that is $\Lambda$
in Eq.~\eqref{eq:UV}, which is equal to $0.560$ GeV in the QM
model, and $0.627$ GeV in the NJL model.

To facilitate the comparison with previous estimates, we review
briefly the frameworks in which the results in Table I are
obtained. In~\cite{Vainshtein:2002nv} the following result is
found, within OPE combined with Pion Dominance (we follow when
possible the notation used in~\cite{Buividovich:2009ih}):
\begin{equation}
\chi^{PD} = -c_\chi\frac{N_c}{8\pi^2 F_\pi^2}~,~~~{\text{Pion
Dominance}} \label{eq:PD}
\end{equation}
with $F_\pi =\sqrt{2}f_\pi= 130.7$ MeV and $c_\chi = 2$; the
estimate of~\cite{Vainshtein:2002nv} is done at a renormalization
point $M = 0.5$ GeV. It is remarkable that Eq.~\eqref{eq:PD} has
been reproduced recently within AdS/QCD approach
in~\cite{Son:2010vc}. Probably, this is the result more comparable
to our estimate, because the reference scales
in~\cite{Vainshtein:2002nv} and in this article are very close.
Within our model calculations we find $c_\chi^{NJL} = 1.93$ and
$c_\chi^{QM} = 2.36$. Using the numerical value of $F_\pi$ and
$c_\chi$ we get $\chi^{PD} = -4.45$ GeV$^{-2}$, which agrees
within $3\%$ with our NJL model result, and within $18\%$ with our
QM model result.

In~\cite{Gorsky:2009ma} the authors find $c_\chi = 2.15$ within
hard-wall holographic approach, at the scale $M \ll 1$ GeV. The
results of~\cite{Gorsky:2009ma} are thus in very good parametric
agreement with~\cite{Vainshtein:2002nv}; on the other hand, the
numerical value of $F_\pi$ in the holographic model is smaller
than the one used in~\cite{Vainshtein:2002nv}, pushing the
holographic prediction for $\chi$ to slightly higher values than
in~\cite{Vainshtein:2002nv}. However, the scale at which the
result of~\cite{Gorsky:2009ma} is valid should be much smaller
than $M=1$ GeV, thus some quantitative disagreement
with~\cite{Vainshtein:2002nv} is expected. As the authors have
explained, it might be possible to tune the parameters of the
holographic model, mainly the chiral condensate, to reproduce the
correct value of $F_\pi$; their numerical tests suggest that by
changing the ratio $\langle\bar ff\rangle/m_\rho$ of a factor of
$8$, then the numerical value of $c_\chi$ is influenced only by a
$5\%$. It is therefore plausible that a best tuning makes the
quantitative prediction of~\cite{Gorsky:2009ma} much closer to the
estimate of~\cite{Vainshtein:2002nv}.

In~\cite{Kim:2004hd} an estimate of $\chi$ within the instanton
vacuum model has been performed beyond the chiral limit, both for
light and for strange quarks (the result quoted in Table I
corresponds to the light quarks; for the strange quark,
$\chi_s/\chi_{u,d} \approx 0.15$ is found). Taking into account
the numerical value of the chiral condensate in the instanton
vacuum, the numerical estimate of~\cite{Kim:2004hd} leads to $\chi
= -2.5\pm 0.15$ GeV$^{-2}$ at the scale $M=1$ GeV. An analytic
estimate within a similar framework has been obtained
in~\cite{Ioffe:2009yi}, in which the zero-mode of the Dirac
operator in the background of a $SU(2)$ instanton is used to
compute the relevant expectation values. The result
of~\cite{Ioffe:1983ju} gives $\chi = -3.52$ GeV$^{-2}$ at
$M\approx 1$ GeV.

In~\cite{Buividovich:2009ih} the result $\chi = -1.547$ GeV$^{-2}$
is achieved within a two-color simulation with quenched fermions.
It is interesting that in~\cite{Buividovich:2009ih} the same
quantity has been computed also at finite temperature in the
confinement phase, at $T=0.82 T_c$, and the result seems to be
independent on temperature. The reference scale
of~\cite{Buividovich:2009ih}, determined by the inverse lattice
spacing, is $M\approx 2$ GeV. Therefore the lattice results are
not quantitatively comparable with our model calculation. However,
they share an important feature with the results presented here,
namely the saturation of the polarization at large values of the
magnetic field. Finally, estimates of the magnetic susceptibility
of the chiral condensate by means of several QCD sum rules there
exist~\cite{Ioffe:1983ju,Belyaev:1984ic,Balitsky:1985aq,Ball:2002ps,Rohrwild:2007yt}.
The results are collected in Table I.

\begin{table*}[t!]
\label{aggiungi}\centering %
\begin{tabular}{|c|c|c|c|}
\hline\hline
{\bf Method} & $\bm\chi$ (GeV$^{-2}$)& {\bf Ren. Point} (GeV)&{\bf Ref.}
\\\hline
\hline Sum rules & $-8.6\pm 0.24$ &1&~\cite{Ioffe:1983ju} \\
\hline Sum rules &-5.7&0.5&~\cite{Belyaev:1984ic} \\
\hline Sum rules & $-4.4\pm0.4$ & 1 &~\cite{Balitsky:1985aq}\\
\hline Sum rules &$-3.15\pm0.3$&1&~\cite{Ball:2002ps} \\
\hline Sum rules & $-2.85\pm0.5$ & 1 &~\cite{Rohrwild:2007yt}\\
\hline OPE + Pion Dominance & $-N_c/(4\pi^2 F_\pi^2)$  &0.5&~\cite{Vainshtein:2002nv} \\
\hline Holography & $-1.075 N_c/(4\pi^2 F_\pi^2)$  & $\ll 1$&~\cite{Gorsky:2009ma} \\
\hline Holography & $-N_c/(4\pi^2 F_\pi^2)$  & $\ll 1$&~\cite{Son:2010vc} \\
\hline Instanton vacuum &$-2.5\pm 0.15$ &1&~\cite{Kim:2004hd} \\
\hline Zero mode of Dirac Operator &-3.52&1&~\cite{Ioffe:2009yi} \\
\hline Lattice&$-1.547(3)$ &2&~\cite{Buividovich:2009ih} \\
\hline NJL model&-4.3&0.63&This work \\
\hline QM model &-5.25&0.56&This work \\
\hline
\end{tabular}
\caption{Magnetic susceptibility of the quark
condensate obtained within several theoretical approaches. In the
table, $F_\pi = 130.7$ MeV. See the text for more details.}
\end{table*}

\section{Renormalized QM model}
In this Section, we make semi-analytic estimates of the
polarization and the magnetic susceptibility of the quark
condensate, as well as for the chiral condensate in magnetic
background, within the renormalized QM model. This is done with
the scope to compare the predictions of the renormalized model
with those of the effective models, in which an ultraviolet cutoff
is introduced to mimic the QCD effective quark mass.

In the renormalized model, we allow the effective quark mass to be
a constant in the whole range of momenta, which is different from
what happens in QCD~\cite{Politzer:1976tv}. Thus, the higher
Landau levels give a finite contribution to the vacuum chiral
condensate even at very strong fields. This is easy to understand:
the ultraviolet cutoff, $\Lambda$, in the renormalized model can
be taken larger than any other mass scale, in particular $\Lambda
\gg |eB|^{1/2}$; as a consequence, the condition $p_3^2 + 2 n |eB|
< \Lambda^2$ is satisfied taking into account many Landau levels
even at very large $eB$. The contribution of the higher Landau
levels, once renormalized, appears in the physical quantities to
which we are interested here, in particular in the chiral
condensate.

Since the computation is a little bit lengthy, it is useful to
anticipate its several steps: firstly we perform regularization,
and then renormalization, of the QEP at zero magnetic field (the
corrections due to the magnetic field turn out to be free of
ultraviolet divergences). Secondly, we solve analytically the gap
equation for the $\sigma$ condensate in the limit of weak fields,
and semi-analytically in the opposite limit. The field-induced
corrections to the QEP and to the solution of the gap equation are
divergence-free in agreement with~\cite{Suganuma:1990nn}, and are
therefore independent on the renormalization scheme adopted. Then,
we compute the renormalized and self-consistent values of the
chiral condensate and of the magnetic moment, as a function of
$eB$, using the results for the gap equation. Within this
theoretical framework, it is much more convenient to compute
$\langle\bar ff\rangle$ and $\Sigma_f$ by taking derivatives of
the renormalized potential; in fact, the computation of the traces
of the propagator in the renormalized model is much more involved
if compared to the situation of the non-renormalized models, since
in the former a non-perturbative (and non-trivial) renormalization
procedure of composite local operators is
required~\cite{Collins:1984xc}. Finally, we estimate $\chi$, as
well as the behavior of the polarization as a function of $eB$.


\subsection{Renormalization of the QEP}

To begin with, we need to regularize the one-loop fermion
contribution in Eq.~\eqref{eq:QEA} namely
\begin{eqnarray}
V_{1-\text{loop}}^\text{fermion} &=& - N_c\sum_f\frac{|Q_f
eB|}{2\pi}\nonumber\\
&&\times\sum_{n=0}^\infty\beta_n
\int_{-\infty}^{+\infty}\frac{dk}{2\pi} \left(k^2 + 2n|Q_f eB| +
m_q^2\right)^{1/2}~.\nonumber\\
&&
\end{eqnarray}
To this end, we define the function, ${\cal V}(s)$, of a complex
variable, $s$, as
\begin{eqnarray}
{\cal V}(s) &=& -N_c\sum_f\frac{|Q_f
eB|}{2\pi}\nonumber\\
&&\times\sum_{n=0}^\infty\beta_n
\int_{-\infty}^{+\infty}\frac{dk}{2\pi} \left(k^2 + 2n|Q_f eB| +
m_q^2\right)^\frac{1-s}{2}~. \nonumber\\
&& \label{eq:Vfs}
\end{eqnarray}

The function ${\cal V}(s)$ can be analytically continued to $s=0$.
We define then $V^{\text{fermion}}_{1-\text{loop}} =
\lim_{s\rightarrow 0^+} {\cal V}(s)$. After elementary integration
over $k$, summation over $n$ and taking the limit $s\rightarrow
0^+$, we obtain the result
\begin{eqnarray}
V_{1-\text{loop}}^\text{fermion} &=& N_c\sum_f\frac{(Q_f
eB)^2}{4\pi^2}\left(\frac{2}{s} -
\log(2|Q_f eB|)+a\right) B_2(q) \nonumber\\
&&-
N_c\sum_f\frac{(Q_f eB)^2}{2\pi^2}\zeta^\prime\left(-1,q\right) \nonumber\\
&&-N_c\sum_f\frac{|Q_f eB| m_q^2}{8\pi^2}\left(\frac{2}{s} -
\log(m_q^2)+a\right)~, \label{eq:F4}
\end{eqnarray}
where we have subtracted terms which do not depend explicitly on
the condensate. In the above equation, $\zeta\left(t,q\right)$ is
the Hurwitz zeta function; for $\text{Re}(t) > 1$ and
$\text{Re}(q) > 0$, it is defined by the series
$\zeta\left(t,q\right) = \sum_{n=0}^\infty(n+q)^{-t}$; the series
can be analytically continued to a meromorphic function defined in
the complex plane $t \neq 1$. Moreover we have defined $q = (m_q^2
+ 2|Q_f eB|)/2|Q_f eB|$; furthermore, $a = 1 - \gamma_E -
\psi(-1/2)$, where $\gamma_E$ is the Eulero-Mascheroni number and
$\psi$ is the digamma function. The derivative $\zeta^\prime
\left(-1,q\right) =d\zeta(t,q)/d t$ is understood to be computed
at $t=-1$.

The first two addenda in Eq.~\eqref{eq:F4} arise from the higher
Landau levels; on the other hand, the last addendum is the
contribution of the LLL. The function $B_2$ is the second
Bernoulli polynomial; using its explicit form, it is easy to show
that the divergence in the LLL term in Eq.~\eqref{eq:F4} is
canceled by the analogous divergence in the first addendum of the
same equation. It is interesting that the LLL contribution, which
is in principle divergent, combines with a part of the
contribution of the higher Landau levels, leading to a finite
result. This can be interpreted as a renormalization of the LLL
contribution. On the other hand, the remaining part arising from
the higher Landau levels is still divergent; this divergence
survives in the $\bm B\rightarrow 0$ limit, and is due to the
usual divergence of the vacuum contribution. We then have
\begin{eqnarray}
V^{\text{fermion}}_{1-\text{loop}} &=&
N_c\sum_f\frac{m_q^4}{16\pi^2}\left(\frac{2}{s}-\log(2|Q_f eB|)+a\right)\nonumber\\
&&+N_c\sum_f\frac{|Q_f eB| m_q^2}{8\pi^2}\log\frac{m_q^2}{2|Q_f
eB|} \nonumber
\\
&&-N_c\sum_f \frac{(Q_f
eB)^2}{2\pi^2}\zeta^\prime\left(-1,q\right)~. \label{eq:F4bis}
\end{eqnarray}

The divergence of the vacuum energy is made explicit in the
expression in the square brackets in Eq.~\eqref{eq:F4bis}. The
mathematical structure of the divergence, namely a pole in $s=0$
and a logarithm with a dimensional argument, is similar to that
obtained within the dimensional regularization scheme. The scale
of the logarithm is hidden in the $1/s$ term, and appears
explicitly when the divergence is subtracted. Such a divergence
affects only the $\bm B=0$ effective potential; the corrections
due to the magnetic field are either finite or independent on the
condensate. As a matter of fact, in the zero magnetic field limit
$q\rightarrow\infty$, the fermion bubble becomes
\begin{equation}
N_c N_f \frac{m_q^4}{16\pi^2}\left(\frac{2}{s}-\log m_q^2+a +
\frac{1}{2}\right)\equiv V_0~, \label{eq:las}
\end{equation}
where we have used the relation
\begin{equation}
\zeta^\prime(-1,q) \approx \left(\frac{1}{12}-\frac{q^2}{4}\right)
+ \log(q)\frac{B_2(q)}{2}~, \label{eq:ZT}
\end{equation}
with $B_2$ corresponding to the second Bernoulli polynomial. Using
Eqs.~\eqref{eq:las} and~\eqref{eq:F4bis} we notice that the pole
$2/s$ is cancelled in the difference $V_1 \equiv
V^{\text{fermion}}_{1-\text{loop}} - V_0$,
\begin{eqnarray}
V_1 &=& -N_c\sum_f\left(\frac{m_q^4}{16\pi^2} + \frac{|Q_f eB|
m_q^2}{8\pi^2}\right)\log\frac{2|Q_f eB|}{m_q^2} \nonumber \\
&&-N_c\sum_f\frac{|Q_f eB|^2}{2\pi^2}\zeta^\prime\left(-1,q\right)
-N_c N_f\frac{m_q^4}{32\pi^2}~. \label{eq:uq}
\end{eqnarray}

To remove the divergence of the vacuum term, we follow
Ref.~\cite{Suganuma:1990nn} and adopt the renormalization
conditions that at zero magnetic field, the quantum corrections do
not shift the classical expectation value of the $\sigma$ field in
the vacuum, $\langle\sigma\rangle=f_\pi$, as well as the classical
value $m_\sigma^2$. To fulfill these conditions, two counterterms
have to be added to the effective potential, $V^{\text{c.t.}} =
\delta\lambda \times \sigma^4/4 + \delta v \times\sigma^2/2$. This
amounts to require
\begin{equation}
\left.\frac{\partial (V_0 +
V^{\text{c.t.}})}{\partial\sigma}\right|_{\sigma = f_\pi} =
\left.\frac{\partial^2 (V_0 +
V^{\text{c.t.}})}{\partial\sigma^2}\right|_{\sigma = f_\pi} = 0~.
\label{eq:RC}
\end{equation}

Taking into account the one-loop divergent contribution and the
condition~\eqref{eq:RC}, we can write the renormalized potential
at zero field as
\begin{eqnarray}
V &=& \frac{\tilde\lambda}{4}\left(\sigma^2-\tilde v^2\right)^2 -
h\sigma - \frac{N_c N_f m_q^4}{16\pi^2}\log\frac{m_q^2}{g^2
f_\pi^2}~, \label{eq:Papapa}
\end{eqnarray}
where $\tilde\lambda$ and $\tilde v$ are the renormalized
parameters, whose expression is not needed here.
Equation~\eqref{eq:Papapa} is in agreement with the textbook
result for the one-loop effective potential of a linear sigma
model coupled to fermions by means of a Yuwaka-type
interaction~\cite{Weinb2}. This is a consistency check of our
regularization and renormalization procedures.

\subsection{Approximate solutions of the gap equation}
{\em Weak fields.} We now restore the magnetic field, and firstly
take the weak field limit. In this context, small field means $eB
\ll m_q^2$. Using again Eq.~\eqref{eq:ZT} we notice that the
derivative of the $\zeta-$function cancels the other addenda in
Eq.~\eqref{eq:uq}, and the remaining contribution is
\begin{eqnarray}
V_1&\approx& - N_c\sum_f\frac{(Q_f
eB)^2}{24\pi^2}\log\frac{m_q^2}{2|Q_f e B|} \nonumber \\
&=& - N_c\sum_f\frac{(Q_f eB)^2}{24\pi^2}\log\frac{m_q^2}{\mu^2}
~, \label{eq:cotta}
\end{eqnarray}
which is in agreement with the result of~\cite{Suganuma:1990nn}.
In the above equation we have followed the notation
of~\cite{Cohen:2008bk} introducing an infrared scale $\mu$,
isolating and then subtracting the term which does not depend on
the condensate. The scale $\mu$ is arbitrary, and we cannot
determine it from first principles; on the other hand, it is
irrelevant for the determination of the $\sigma-$condensate. We
expect $\mu\approx f_\pi$ since this is the typical scale of
chiral symmetry breaking in the model for the $\sigma$ field. The
correction~\eqref{eq:cotta} lowers the effective potential of the
broken phase, thus favoring the spontaneous breaking of chiral
symmetry. The result in Eq.~\eqref{eq:cotta} is UV divergence-free
as anticipated; thus, it is independent on the particular
procedure used to regularize the QEP.

In this limit, it is easy to obtain analytically the behavior of
the constituent quark mass as a function of $eB$. As a matter of
fact, we can expand the derivative of the QEP with respect to
$\sigma$, around the solution at $B=0$, writing
$\langle\sigma\rangle = f_\pi + \delta\sigma$. Then, a
straightforward evaluation leads to
\begin{equation}
m_q = g f_\pi\left( 1 + \frac{5}{9}\frac{N_c}{12\pi^2 f_\pi^2
m_\sigma^2}  (eB)^2 \right)~. \label{eq:WEA}
\end{equation}
As anticipated, the scale $\mu$ is absent in the solution of the
gap equation.

{\em Strong fields.} In the limit $eB\gg m_q^2$, we can find an
asymptotic representation of $V_1$ by using the expansion
$\zeta^\prime(-1,q) = c_0 + c_1 (q-1)$ valid for $q\approx 1$,
with $c_0 = -0.17$ and $c_1 = -0.42$. Then we find
\begin{eqnarray}
V_1 &\approx& -N_c\sum_f \frac{m^2_q}{8\pi^2}\left(\frac{m^2_q}{2}
+ |Q_f eB|\right)\log\frac{2|Q_f eB|}{m^2_q} \nonumber
\\&& - N_c\sum_f\frac{|Q_f eB|m_q^2}{2\pi^2}c_1~, \label{eq:Ft}
\end{eqnarray}
where we have subtracted condensate-independent terms.

In the strong field limit it is not easy to find analytically an
asymptotic representation for the sigma condensate as a function
of $eB$; therefore we solve the gap equation numerically, and then
fit data with a convenient analytic form as follows:
\begin{equation}
m_q = b|eB|^{1/2} + \frac{c f_\pi^3}{|eB|}~, \label{eq:SEA}
\end{equation}
where $b=0.32$ and $c=32.78$. At large fields the quark mass grows
as $|eB|^{1/2}$ as expected by dimensional analysis; this is a
check of the equations that we use.

\subsection{Evaluation of chiral condensate and magnetic moment}
{\em Chiral condensate.} To compute the chiral condensate we
follow a standard procedure: we introduce source term for $\bar
ff$, namely a bare quark mass $m_f$, then take derivative of the
effective potential with respect to $m_f$ evaluated at $m_f = 0$.
This amounts to derive only $V^{\text{fermion}}_{1-\text{loop}}$,
since both the classic potential and counterterms do not have a
dependence on $m_f$. According to Eq.~\eqref{eq:uq} we find
\begin{equation}
\langle\bar ff\rangle = \left.\frac{\partial V_0}{\partial
m_f}\right|_{m_f = 0} + \left.\frac{\partial V_1}{\partial
m_f}\right|_{m_f = 0}~.\label{eq:hhh}
\end{equation}
The first and second addenda on the r.h.s. of Eq.~\eqref{eq:hhh}
represent the vacuum and the field-induced contributions to the
chiral condensate respectively. The vacuum contribution is
divergent, and can be renormalized according to the procedure of
renormalization of composite local
operators~\cite{Collins:1984xc}. It is not necessary to perform
this procedure here, because the numerical value of the vacuum
condensate is not necessary in our discussion\footnote{In
principle, in the renormalization program of the composite
operator $\langle\bar qq\rangle$ one can impose, as a
renormalization condition, that the chiral condensate in the
vacuum is consistent with the results of~\cite{Dosch:1997wb} at
the renormalization point $M=1$ GeV.}. Therefore, it is enough to
compute only the contribution at $\bm B \neq 0$ arising from
$V_1$, which is finite.

In particular, for the weak field case we obtain
\begin{eqnarray}
\langle\bar ff\rangle &=& \langle\bar ff\rangle_0 -
\frac{N_c}{12\pi^2}\frac{ |Q_f e B|^2}{m_q}~. \label{eq:CCggg}
\end{eqnarray}
On the other hand, in the strong field limit we have
\begin{equation}
\langle\bar ff\rangle = -\frac{N_c m_q}{4\pi^2} \left(|Q_f e B| +
m_q^2\right)\log\frac{2|Q_f e B|}{m_q^2} ~. \label{eq:CCsss}
\end{equation}
Using Equations~\eqref{eq:WEA} and~\eqref{eq:SEA}, we show that
the chiral condensate scales as $a + b (eB)^2$ for small fields,
and as $|eB|^{3/2}$ for large fields.

{\em Magnetic moment.} Next we turn to the computation of the
expectation value of the magnetic moment. The expression in terms
of Landau levels is given by Eq.~\eqref{eq:TrM}, which clearly
shows that this quantity has a log-type divergence. In order to
avoid a complicated renormalization procedure of a local composite
operator, we notice that it is enough to take the minus derivative
of $V_1$ with respect to $\bm B$ to get magnetization, ${\cal
M}$~\cite{Cohen:2008bk}, then multiply by $2m/Q_f$ to get the
magnetic moment. This procedure is very cheap, since the $\bm
B-$dependent contributions to the effective potential are finite,
and the resulting expectation value will turn out to be finite as
well (that is, already renormalized).

In the case of weak fields, from Eq.~\eqref{eq:cotta} we find
\begin{equation}
\Sigma_f = N_c \frac{Q_f |eB|
m_q}{6\pi^2}\log\frac{m_q^2}{\mu^2}~. \label{eq:WFmm}
\end{equation}
On the other hand, in the strong field limit we get from
Eq.~\eqref{eq:Ft}
\begin{equation}
\Sigma_f = N_c \frac{m_q^3}{4\pi^2}\log\frac{2|Q_f eB|}{m_q^2}~.
\label{eq:SFmm}
\end{equation}
The above result is in parametric agreement with the estimate of
magnetization in~\cite{Cohen:2008bk}. In fact, $m_q^2\approx |eB|$
in the strong field limit, which leads to a magnetization ${\cal
M}\approx B\log B$.

Using the expansions for the sigma condensate at small and large
values of the magnetic field strength, we argue that $\Sigma_f
\approx |eB|$ in a weak field, and $\Sigma_f \approx |eB|^{3/2}$
in a strong field.

\subsection{Computation of chiral magnetization and polarization}
\begin{figure}[t!]
\begin{center}
\includegraphics[width=8.5cm]{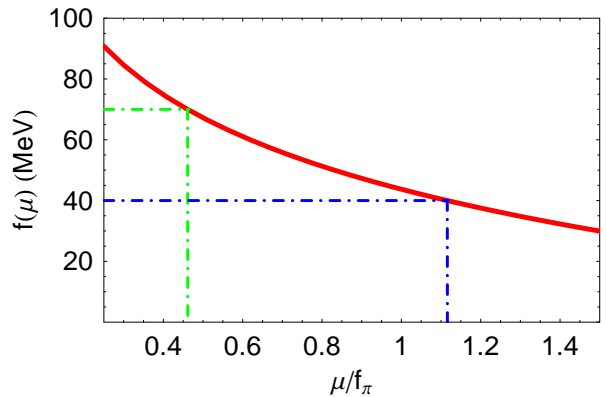}
\end{center}
\caption{\label{Fig:4} Magnetic susceptibility of the quark
condensate multiplied by the chiral condensate, in MeV, as a
function of the infrared scale $\mu$, as given by
Eq.~\eqref{eq:slope}. The interval on the $\mu$-axis delimited by
the green and the blue vertical lines is the range in which we
obtain a value of $\chi$ which is consistent with phenomenology.}
\end{figure}

We can now estimate the magnetic susceptibility of the quark
condensate and the polarization as a function of $eB$. For the
former, we need to know the behavior of the magnetic moment for
weak fields. From Eq.~\eqref{eq:WFmm} and from the
definition~\eqref{eq:chiDef} we read
\begin{equation}
\chi\langle\bar ff\rangle =\frac{N_c
m_q}{6\pi^2}\log\frac{m_q^2}{\mu^2}\equiv f(\mu)~.
\label{eq:slope}
\end{equation}
The presence of the infrared scale $\mu$ makes the numerical
estimate of $\chi$ uncertain; however, taking for it a value
$\mu\approx f_\pi$, which is the typical scale of chiral symmetry
breaking, we have $\chi\langle\bar ff\rangle \approx 44$ MeV,
which is in agreement with the expected value, see
Eq.~\eqref{eq:i1}. In Fig.~\ref{Fig:4} we plot $f(\mu)$ as a
function of $\mu$. The interval on the $\mu$-axis delimited by the
green and the blue vertical lines is the range in which we obtain
a value of $\chi$ which is consistent with phenomenology.

Next we turn to the polarization. For weak fields we find
trivially a linear dependence of $\mu_f$ on $|Q_f eB|$, with slope
given by the absolute value of $\chi$ in Eq.~\eqref{eq:slope}. On
the other hand, in the strong field limit we find, according to
Eq.~\eqref{eq:SFmm},
\begin{equation}
\mu_f \approx \frac{m_q^2}{m_q^2 + |Q_f eB|} \approx 1 -
\frac{|Q_f|}{b + |Q_f|}~, \label{eq:asymSF}
\end{equation}
where we have used Eq.~\eqref{eq:SEA}. This result shows that the
polarization saturates at large values of $eB$, but the asymptotic
value depends on the flavor charge.

It is interesting to compare the result of the renormalized model
with that of the effective models considered in the previous
Section. In the former, the asymptotic value of $\mu_f$ is
flavor-dependent; in the latter, $\mu_f \rightarrow 1$
independently on the value of the electric charge. Our
interpretation of this difference is as follows: comparing
Eq.~\eqref{eq:asymSF} with the general model expectation,
Eq.~\eqref{eq:HLL}, we recognize in the factor $|Q_f|/(b+|Q_f|)$
the contribution of the higher Landau levels at zero temperature,
which turns out to be finite and non-zero after the
renormalization procedure. This contribution is then transmitted
to the physical quantities that we have computed. The trace of the
higher Landau levels is implicit in the solution of the gap
equation in the strong field limit, namely the factor $b$ in
Eq.~\eqref{eq:SEA}, and explicit in the additional $|Q_f|$
dependence in Eq.~\eqref{eq:asymSF}. A posteriori, this conclusion
seems quite natural, because in the renormalization procedure we
assume that the effective quark mass is independent on quark
momentum, thus there is no cut of the large momenta in the gap
equation (and in the equation for polarization as well). In the
effective models considered in the first part of this article, on
the other hand, the cutoff procedure is equivalent to have a
momentum-dependent effective quark mass, $m_q =
g\sigma\Theta(\Lambda^2 - p_3^2 - 2n|Q_f eB|)$, which naturally
cuts off higher Landau levels when $eB \gg \Lambda^2$. At the end
of the days, the expulsion of the higher Landau levels from the
chiral condensate makes $\mu_f \rightarrow 1$ in the strong field
limit. Our expectation is that if we allow the quark mass to run
with momentum and decay rapidly at large momenta, mimicking the
effective quark mass of QCD, higher Landau levels would be
suppressed in the strong field limit, and the
result~\eqref{eq:asymSF} would tend to the result in
Fig.~\ref{Fig:2}.

\section{Conclusions}
In this article, we have computed the magnetic susceptibility of
the quark condensate, $\chi$, and the polarization, $\mu_f$, in a
background of a magnetic field, $\bm B$, by means of two chiral
models of QCD: the quark-meson model and the Nambu-Jona-Lasinio
model. The knowledge of these quantities is relevant both
theoretically and phenomenologically. Indeed, the magnetic
susceptibility of the quark condensate might lead, in the
photoproduction of lepton pairs, to an interference between the
Drell-Yan process and the photon-quark coupling, the latter
induced by the chiral-odd distrubution amplitude of the
quark~\cite{Pire:2009ap}. The two models are widely used to study
the phase diagram of QCD in several regimes; it has been proved in
different contexts, that they offer a good theoretical tool to
compute low-energy QCD properties. It is thus interesting to
compute, within these models, quantities which characterize the
QCD vacuum in a magnetic background, and compare the results with
those obtained within different theoretical frameworks, both
analytically and numerically.

In the first part of this article, we have reported our results
about $\chi$ and $\mu_f$ obtained within a numerical
self-consistent solution of the model, in the one-loop
approximation. Our results on $\chi$ are summarized in Table~I,
and are in fair agreement with previous estimates. Besides, our
data on polarization are collected in Fig.~3. For the latter, we
obtain a saturation to the asymptotic value $\mu_\infty = 1$ at
large values of $eB$, which is understood as a lowest Landau level
dominance in the quark condensate. This is in agreement with
recent results obtained within Lattice QCD simulations with two
colors and quenched fermions~\cite{Buividovich:2009ih}.

In the second part of the article, we have estimated $\chi$ and
$\mu_f$ within a renormalized version of the QM model. The
numerical value of $\chi$ in this case is quite uncertain because
of the presence, in the final result, of an unknown infrared scale
$\mu$. However, as shown in Fig.~\ref{Fig:4}, taking for $\mu$ a
numerical value around $f_\pi$, which is the typical scale of
chiral symmetry breaking in this model, we obtain a value of
$\chi$ which is consistent with phenomenology.

Within the renormalized model we find a saturation of $\mu_f$ at
large $\bm B$, in qualitative agreement with our findings within
the effective models in Section III and quenched
QCD~\cite{Buividovich:2009ih}. In the renormalized model, the
asymptotic value of $\mu_f$ is flavor-dependent; in the effective
models and quenched QCD, $\mu_f \rightarrow 1$ independently on
the value of the electric charge. We attribute this difference to
the presence, in the theory, of the renormalized contribution of
the higher Landau levels: in the renormalized model, the higher
Landau levels give a finite contribution to the chiral condensate
at zero temperature, even in the case of very strong fields. After
renormalization of the QEP, this contribution is finite and
non-zero, and is transmitted to the physical quantities that we
have computed.

The results we obtained are quite encouraging, and suggest that a
systematic study of external field induced condensates at zero, as
well as finite, temperature within chiral models is worth to be
done. As a natural continuation of this study, it would be
interesting to add the strange quark, following the
work~\cite{Kim:2004hd}. Moreover, Lattice simulations have shown
that $\chi$ is almost insensitive to the temperature, at least in
the confinement phase~\cite{Buividovich:2009ih}. This result is
achieved with quenched fermions. On the other hand, the chiral
model handle with dynamical fermions; it would be therefore of
interest to address the question of the behavior of $\chi$ as a
function of temperature using chiral models. Furthermore, it has
been shown~\cite{Buividovich:2009my} that local fluctuations of
topological charge induce a quark electric dipole moment along the
direction of a strong magnetic field. This problem can be studied
easily within the chiral models, introducing a pseudo-chemical
potential $\mu_5$ conjugated to chirality imbalance as already
done in~\cite{Fukushima:2010fe,Chernodub:2011fr}.

{\bf Acknowledgements}. We acknowledge M.~Chernodub, T.~Cohen, A.
Flachi, R.~Gatto, A.~Ohnishi for several discussions and
correspondence. Moreover, we thank H. Suganuma for valuable and
numerous discussions on the renormalization of the quark-meson
model and for a careful reading of the manuscript. The Laboratoire
de Matematique et Physique Teorique of Tours University, as well
as the Physics Department of Jyvaskyla University, where part of
this work was completed, are acknowledged for the kind
hospitality. The work of M.~R. is supported by JSPS under contract
number P09028.

\end{document}